\documentclass[prl,twocolumn,amsmath,amssymb,floatfix]{revtex4}
\usepackage{graphicx}

\begin{document}
\graphicspath{{FiguresForPaper/}}

\title{A Superconducting ``Dripping Faucet''}

\author{Stuart B. Field and Gheorghe Stan}
\affiliation {Department of Physics, Colorado State University, Fort Collins, CO 80523}

\begin{abstract}
When a current is applied to a type-I superconducting strip containing a narrow channel across its width, magnetic flux spots nucleate at the edge and are then driven along the channel by the current. These flux ``drops'' are reminiscent of water drops dripping from a faucet, a model system for studying low-dimensional chaos. We use a novel high-bandwidth Hall probe to detect in real time the motion of individual flux spots moving along the channel. Analyzing the time series consisting of the intervals between successive flux drops, we find distinct regions of chaotic behavior characterized by positive Lyapunov exponents, indicating that there is a close analogy between the  {\it dynamics} of the superconducting and water drop systems. 
\end{abstract}

\maketitle

Water dripping from a faucet is an everyday phenomenon that illustrates many of the basic ideas of deterministic chaos \cite{Martien85,Wu89a,Dreyer91}.  As the drip rate is increased, the intervals between successive drops pass through regimes of periodic, multiply-periodic, and chaotic behavior. In the chaotic regime, the drop dynamics is deterministic, in that the next several drop intervals can be predicted from previous intervals. There is, however,  no {\it long-term} predictability of the system; a small change in an earlier interval will lead to an exponentially different set of future intervals. In a simple model \cite{Shaw:1984}, a drop grows until it reaches a threshold size; it then detaches and falls under the influence of gravity.  The remaining water oscillates as a drop begins to grow again.  A key element of this model is that the time that any drop detaches is {\it causally related} to the time the previous drop detached, via the ``memory'' of that previous time stored in the oscillations.  It is in this way that the time interval between two drops is deterministically related to earlier drop intervals. Yet, because of the delicacy of the threshold condition, small changes in the state of an earlier drop can have a profound effect on later ones.

Interestingly, very similar physics may govern the current-driven nucleation, growth, and breakoff of magnetic {\it flux spots} or {\it drops} in type-I superconductors. When current is passed along a thin-film strip of type-I material containing a narrow channel across its width, flux begins to enter the channel at its ends  \cite{Chimenti:1976,Chimenti:1977,Chimenti:1978,Hurm:1979,Muhlemeier:1986}. This region of flux grows to some critical size and then breaks off as a flux spot, containing $\sim\!100$ flux quanta $\Phi_0$, which is then driven down the channel by the current. The analogy with water drops is readily apparent, and suggests that in this {\it superconducting dripping faucet} chaotic dynamics might be observable for flux drops as well.

In the experiment we report here, a high-bandwidth Hall sensor is used to directly measure the passage of individual flux spots along a channel formed in a lead strip. Just as for water drops falling from a faucet, we find that the flux spot dynamics can be periodic, with single or multiple periods, or nonperiodic, with a broad distribution of drop intervals. Applying the tools of nonlinear time-series analysis to the data, we find that these nonperiodic regimes are in fact chaotic, characterized by positive Lyapunov exponents, allowing the prediction of the drop sequence roughly five drops into the future.

\begin{figure}
\includegraphics[width=2.8in]{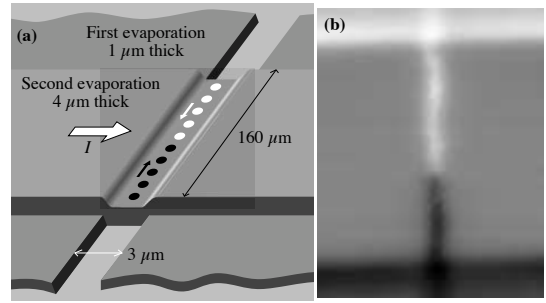}
\caption{(a) Schematic of the lead sample, showing the narrow channel that confines flux spots. Spots of opposite sign nucleate at each end and move towards the middle of the channel, where they annihilate. (b) Scanning Hall probe image of the sample for $I > I_c$. The low bandwidth of the scanning mode leads to a blurred image of the moving spots.
\label{FIG_Sample}}
\end{figure}

The sample geometry used is similar to that of Chimenti {\it et al.}~\cite{Chimenti:1976}. In a two-step process, a 1-$\mu$m-thick lead film is first evaporated onto a sapphire substrate, leading to a 1-mm-wide strip bridged by a 3-$\mu$m-wide gap defined by liftoff. Then a second lead strip, 4~$\mu$m thick and 160~$\mu$m wide, is evaporated on top of the first. As shown schematically in Fig. \ref{FIG_Sample}(a), this results in a channel that is 3~$\mu$m wide, 1~$\mu$m deep, and 160~$\mu$m long. The films are of high quality, with $T_{\rm{c}} \approx 7.2$~K and $R_{\rm{300~K}}/R_{\rm{10~K}} > 500$.

Hall sensors, with active areas of $\approx1.5 \times 1.5$~$\mu$m, were fabricated from high-quality GaAs/AlGaAs heterostructures. The Hall voltage was detected using a cooled JFET preamplifier mounted near the Hall probe.
At room temperature, the Hall signal was further amplified and then passed through an antialiasing filter with a bandwidth of 5 MHz. The Hall probes were mounted on a scanning head that allowed the probe to be moved over any point of the channel, or to be scanned over the surface for magnetic imaging.

In dripping faucet experiments, the drop dynamics are typically investigated as a function of the average drip rate, a rate usually controlled by varying the pressure behind the nozzle. In our experiment, we can vary the flux-spot nucleation rate by changing the external current flowing through the strip. Figure~\ref{FIG_Sample}(b) shows a magnetic image, taken with the Hall probe in scanning mode, of the strip with a current flowing along it. The bright band along the upper edge of the sample, and the dark band along its lower edge, reflect the magnetic field due to this applied current. When the current exceeds a critical value $I_c$, this magnetic field becomes large enough to allow the nucleation and subsequent breakoff of flux drops from the channel ends. Because the field is oppositely directed at the two edges, the flux drops from each edge have opposite signs as well. As Fig.~\ref{FIG_Sample}(b) shows, these drops of opposite sign are then driven by the current toward the middle of the channel, where they annihilate.

In order to detect the motion of individual flux spots in real time, the Hall probe was held fixed over one point of the channel, about 40 $\mu$m (or 1/4 of its length) from the channel's end. Here, we are well away from the very edge where the flux spots nucleate and break off, so that we observe the dynamics of only well-formed flux spots. At the same time, this position is far from the annihilation point, so that we avoid the complicating effects of the annihilation process. 

The results reported here were obtained in zero applied field at a temperature of 4.5 K. At higher temperatures close to $T_c$, the magnetic field of the flux spots is weak and difficult to observe; at low temperatures the critical current becomes impracticably large. The results at other temperatures, or for other samples, look qualitatively similar, but differ in their details. At 4.5 K, the current was swept from the critical current $I_c = 497$~mA to a threshold current $I_t=590$~mA at which continuous flux flow occurred in the channel.  The Hall probe signal was digitized at a rate of $10^7$ samples/s, with a total of $16\times 10^6$ points taken in one run.

\begin{figure}
\includegraphics[width=2.8in]{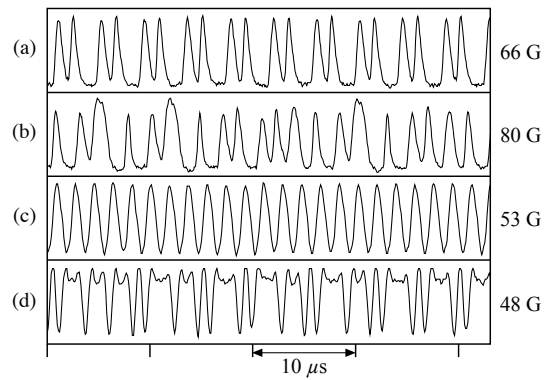}
\caption{Sequences of voltage pulses recorded by the Hall probe at four values of the current $\Delta I = I - I_c$: (a)~12.4~mA; (b)~15.9~mA; (c)~30.1~mA; (d)~81.8~mA. The values to the right are the field scale for each. 
\label{pulses}}
\end{figure}

Figure~\ref{pulses} shows short segments of an entire data run.The signal consists of well-defined Hall voltage pulses as each flux spot passes beneath the probe. In Fig.~\ref{pulses}(a), taken at a current somewhat above the critical current, we see a train of pulses with two distinct periods. As the current is increased, the behavior changes to that shown in Fig.~\ref{pulses}(b). Here we observe a complex train of larger and smaller pulses, with no evident periodicity. As the current is swept through this low-current {\it Region I}, the behavior changes alternately between the periodic type behavior seen in Fig.~\ref{pulses}(a) and the more complex behavior in Fig.~\ref{pulses}(b). A fundamental question to be addressed concerns the nature of the complex dynamical behavior in Fig.~\ref{pulses}(b). Are the pulse times and sizes only the result of some stochastic process, or is the underlying dynamics in fact deterministic?

\begin{figure*}
\includegraphics[width=6.5in]{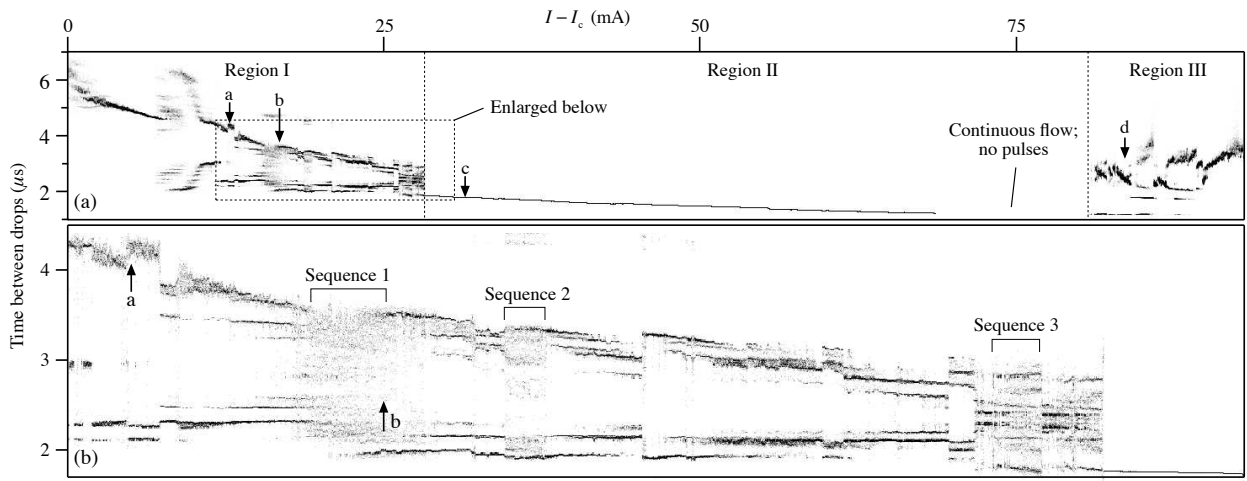}
\caption{Bifurcation diagram for ``dripping'' time intervals. (a) Drop intervals over the entire range of currents for which pulses were observed. Arrows a--d denote the currents at which the segments of pulses shown in Fig.~\ref{pulses} were taken. (b) An expanded section of Regions I and II. Short sequences 1--3 are analyzed in more detail using the tools of nonlinear analysis.
\label{Bifurcation}}
\end{figure*}

As the current is further increased, the behavior changes suddenly to the purely period-one behavior shown in Fig.~\ref{pulses}(c). We call the fairly wide range of currents over which this periodic behavior is observed {\it Region II}. Finally, at the highest currents, the flux-flow dynamics enters {\it Region III}. As shown in Fig.~\ref{pulses}(d), the Hall voltage consists of fairly flat-topped pulses interspersed with occasional short pulses. We interpret this pattern as representing elongated flux ``sausages'' interspersed with more circular flux spots. As the current is further increased, Region III  ends when the flux spots merge and  {\it continuous} flux flow occurs in the channel.  

One run of 16 million data points contains some 600,000 pulses representing a variety of dynamical regimes. In order to analyze this large data set we need a way to characterize the data in a concise way. For water drop experiments a commonly used measure is the time interval $\Delta t$ between successive drops. Each drop interval can then be plotted versus the driving parameter, such as the water pressure at the nozzle. We have found it useful to plot our data in a similar way.  The time at which a flux ``drop'' occurs is determined by when the voltage crosses a certain threshold. The resulting {\it bifurcation diagram} is shown in Fig.~\ref{Bifurcation}(a), in which we plot the time intervals $\Delta t$ between the 638,848 individual drops observed during the run. The gray-scale intensity of the image is proportional to the probability of finding, at a given driving current, a particular value of $\Delta t$.  The three regions I--III previously described are readily apparent. 

Region I is characterized by regions of singly- and multiply-periodic behavior interspersed with more complex regions distinguished by broad distributions of ``dripping" time intervals. We'll discuss this interesting region in more detail below. In Region II, purely periodic behavior is observed, as indicated by the {\it single} $\Delta t$ observed at any given current. As the current is increased, this period decreases, indicating higher-frequency flux nucleation. Finally, at high currents, Region III emerges. This region is similar in appearance to Region I, with some periodic sections mixed with more complex regimes. (At the tail end of Region II there is a short section with continuous flux flow.)

In Fig.~\ref{Bifurcation}(b) is shown an enlarged view of the most interesting section of Region I. At this level of detail it is clear, for example, that the pulse train in Fig.~\ref{pulses}(a) is not truly periodic. First, the shorter drop interval observed in  Fig.~\ref{pulses}(a) can be seen to actually consist of {\it two} possible intervals of 2.11 and 2.25 $\mu$s. Second, the longer time interval of about 4.2 $\mu$s is broadened by about 0.16 $\mu$s. At the current corresponding to the pulse train shown in Fig.~\ref{pulses}(b), a very broad range of drop intervals is apparent, reflecting the multitude of pulse intervals observed in Fig.~\ref{pulses}(b). Other regions in Fig.~\ref{Bifurcation}(b) also exhibit multiple periodicity, quasiperiodicity, and broad distributions of $\Delta t$ with no evident periodicity.

We have discussed how the dynamics of a real dripping faucet is governed by highly nonlinear processes. Our qualitative analysis of our ``superconducting dripping faucet" suggests that nonlinear processes are at play here as well. And, since flux drops represent a driven, dissipative system  it seems possible that the system's behavior is determined by a chaotic attractor rather than just being a stochastic manifestation of flux spot nucleation.  To investigate this we have analyzed the sequence of 638,847 drop intervals using the TISEAN~\cite{TISEAN} package for nonlinear time series analysis, calculating specific quantities such as the largest Lyapunov exponent and the correlation fractal dimension. In order to make such an analysis, stationary sequences---those in which the underlying governing dynamics is unchanging---should be used. We have chosen the finite sequences 1--3 indicated in Fig.~\ref{Bifurcation}(b) as approximations of true stationary sequences. These sequences contain 4000, 2000, and 3000 drop intervals, respectively.
Our quantitative analysis is performed in the associated phase space constructed using the method of time delay reconstruction~\cite{Takens81}, and a nonlinear noise reduction algorithm was applied to the data before further analysis.

\begin{figure}
\includegraphics[width=3.375in]{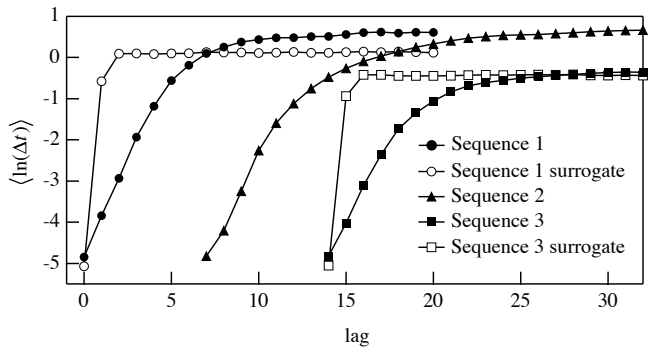}
\caption{Divergences of trajectories from sequences 1--3 of Fig.~\ref{Bifurcation}. The graphs for sequences 2 and 3 are offset horizontally by 7 and 14 units respectively. The embedding dimension for this analysis was $m = 5$.
\label{Lyapunov}}
\end{figure}

We begin our time-series analysis by computing Lyapunov exponents, which characterize the evolution of the separation between two nearby trajectories in phase space. If the dynamics is governed by deterministic chaos then nearby trajectories diverge exponentially and the largest Lyapunov exponent is positive. We have used the algorithm of Kantz~\cite{Kantz94,TISEAN} to study this divergence.  For sequence 1 at the left of Fig.~\ref{Lyapunov}, the average separation between points in phase space, starting with an average separation of about 0.008 ($\ln(0.008) = -4.8$), increases {\it linearly} on this log-lin graph, so that there is indeed an exponential divergence of nearby trajectories. This linear increase extends over about 4--5 consecutive steps, indicating that weak correlations still exist between a given drop interval and one four or five drops later. Similar results hold for sequences 2 and 3.  At large enough time steps the originally nearby points become completely uncorrelated, and the curve begins to approach the size of the attractor. The curves with unfilled markers in Fig.~\ref{Lyapunov} are calculated for surrogate data obtained by phase-randomizing the data \cite{Schreiber00} from sequences 1 and 3. In this case, the slope of the average expansion rate is almost vertical: Any two points are completely uncorrelated, and their average distance immediately jumps to the average size of the (randomized) attractor. This clear distinction between the original and the surrogate data proves that our dynamics is incompatible with a linear stochastic process, but instead is well-described by a nonlinear deterministic process.

\begin{figure}
\includegraphics[width=3.375in]{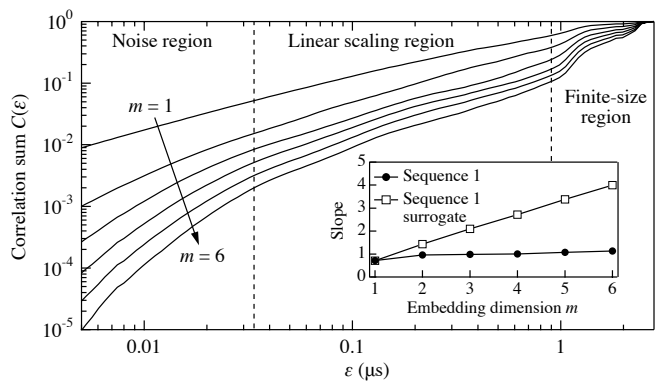}
\caption{The correlation sum computed for sequence 1. For small $\epsilon$ the sum is dominated by noise; at large values the finite size of the attractor becomes important. Over about 1.4 decades, however, scaling is observed. (Inset) Slopes of the correlation sum as measured in the scaling region.
\label{Correlation}}
\end{figure}

The {\it correlation dimension}~\cite{Grassberger83} quantifies the self-similarity exhibited by the attractor's structure in phase space. 
By counting the points inside of a ball which is moved along the phase-space trajectory we obtain the correlation sum $C(\varepsilon)$ as a function of the ball radius $\varepsilon$. In Fig.~\ref{Correlation} we show the phase-space distance-dependence of the correlation sum for sequence 1, calculated for different embedding dimensions \(m\). The linear scaling region between 0.034 \(\mu\)s and 0.09 \(\mu\)s is due to the points in the phase space that are arranged into structures with internal self-similar organization.  The inset to Fig.~\ref{Correlation} shows the embedding-dimension dependence of the slope of the correlation sum's linear region. As the embedding dimension \(m\) is increased, the scaling coefficient approaches the correlation dimension $D \approx 1.0$ of the attractor formed by sequence 1. However, in the randomized surrogate data of sequence 1,  the correlation sum describes only a stochastic distribution of points in phase space, and the scaling coefficient shows no asymptotic behavior.

We have shown that the dynamics of our superconducting dripping faucet is governed by nonlinear dynamics, leading to chaotic behavior very similar to that observed in an ordinary dripping faucet. The question remains as to what extent the two systems share an underlying physical origin for this behavior. A key element of water drop dynamics is the ``memory" that each drop has of its predecessors, stored in the oscillations of the drop itself. But flux motion in a superconductor is heavily overdamped, and so no such oscillations are expected. However, unlike water drops, flux spots {\it interact}, via long-range magnetic \cite{Buck89} forces. Thus an incipient spot's development is mediated by interactions with the spot that just broke off, and even with spots further down the channel. In this way the time intervals between drops is a deterministic---but evidently highly nonlinear---function of previous drops.

Although no general theory exists for calculating this nucleation, growth, and breakoff mechanism, the remarks just given are enough to envision the kinds of correlations that such a theory would yield. At low driving currents, drops are well-separated and, experimentally, their nonlinear interactions are evidently such that the breakoff time of the next drop is sensitively dependent on the position and size of previous drops. This leads to the chaotic behavior observed in Region I. As the current is further increased, the drop intervals shrink and the interactions become stronger. It appears that in this regime the interactions act to ``lock'' together successive drops in a highly periodic way, leading to our observed Region II. At the tail end of Region II the drops are so closely spaced that they merge into {\it continuous} flux flow. At the highest currents, this continuous flow begins to break up again, but as Fig.~\ref{pulses}(d) showed, the signal consists regions of flux separated by short segments of zero flux. We can think of these short segments as {\it negative}-going pulses moving in a ``sea'' of continuous positive flux, in which case the dynamics should be similar to that in Region I, as is in fact observed.

The authors are grateful to Franco Nori for valuable discussions. This research was supported in part by NSF grant DMR-0308699.

\bibliography{./References/References}

\newcommand{\noopsort}[1]{} \newcommand{\printfirst}[2]{#1}
  \newcommand{\singleletter}[1]{#1} \newcommand{\switchargs}[2]{#2#1}
\begin{thebibliography}{15}
\expandafter\ifx\csname natexlab\endcsname\relax\def\natexlab#1{#1}\fi
\expandafter\ifx\csname bibnamefont\endcsname\relax
  \def\bibnamefont#1{#1}\fi
\expandafter\ifx\csname bibfnamefont\endcsname\relax
  \def\bibfnamefont#1{#1}\fi
\expandafter\ifx\csname citenamefont\endcsname\relax
  \def\citenamefont#1{#1}\fi
\expandafter\ifx\csname url\endcsname\relax
  \def\url#1{\texttt{#1}}\fi
\expandafter\ifx\csname urlprefix\endcsname\relax\def\urlprefix{URL }\fi
\providecommand{\bibinfo}[2]{#2}
\providecommand{\eprint}[2][]{\url{#2}}

\bibitem[{\citenamefont{Martien et~al.}(1985)\citenamefont{Martien, Pope,
  Scott, and Shaw}}]{Martien85}
\bibinfo{author}{\bibfnamefont{P.}~\bibnamefont{Martien}},
  \bibinfo{author}{\bibfnamefont{S.~C.} \bibnamefont{Pope}},
  \bibinfo{author}{\bibfnamefont{P.~L.} \bibnamefont{Scott}}, \bibnamefont{and}
  \bibinfo{author}{\bibfnamefont{R.~S.} \bibnamefont{Shaw}},
  \bibinfo{journal}{Phys. Lett. A} \textbf{\bibinfo{volume}{110}},
  \bibinfo{pages}{399} (\bibinfo{year}{1985}).

\bibitem[{\citenamefont{Wu and Schelly}(1989)}]{Wu89a}
\bibinfo{author}{\bibfnamefont{X.}~\bibnamefont{Wu}} \bibnamefont{and}
  \bibinfo{author}{\bibfnamefont{Z.}~\bibnamefont{Schelly}},
  \bibinfo{journal}{Physica D} \textbf{\bibinfo{volume}{40}},
  \bibinfo{pages}{433} (\bibinfo{year}{1989}).

\bibitem[{\citenamefont{Dreyer and Hickey}(1991)}]{Dreyer91}
\bibinfo{author}{\bibfnamefont{K.}~\bibnamefont{Dreyer}} \bibnamefont{and}
  \bibinfo{author}{\bibfnamefont{F.}~\bibnamefont{Hickey}},
  \bibinfo{journal}{Am. J. Phys.} \textbf{\bibinfo{volume}{59}},
  \bibinfo{pages}{619} (\bibinfo{year}{1991}).

\bibitem[{\citenamefont{Shaw}(1984)}]{Shaw:1984}
\bibinfo{author}{\bibfnamefont{R.}~\bibnamefont{Shaw}},
  \emph{\bibinfo{title}{The Dripping Faucet as a Model Chaotic System}}
  (\bibinfo{publisher}{Aerial Press}, \bibinfo{address}{Santa Cruz},
  \bibinfo{year}{1984}).

\bibitem[{\citenamefont{Chimenti et~al.}(1976)\citenamefont{Chimenti, Watson,
  and Huebener}}]{Chimenti:1976}
\bibinfo{author}{\bibfnamefont{D.}~\bibnamefont{Chimenti}},
  \bibinfo{author}{\bibfnamefont{H.}~\bibnamefont{Watson}}, \bibnamefont{and}
  \bibinfo{author}{\bibfnamefont{R.}~\bibnamefont{Huebener}},
  \bibinfo{journal}{J. Low Temp. Phys.} \textbf{\bibinfo{volume}{23}},
  \bibinfo{pages}{303} (\bibinfo{year}{1976}).

\bibitem[{\citenamefont{Chimenti and Huebener}(1977)}]{Chimenti:1977}
\bibinfo{author}{\bibfnamefont{D.}~\bibnamefont{Chimenti}} \bibnamefont{and}
  \bibinfo{author}{\bibfnamefont{R.}~\bibnamefont{Huebener}},
  \bibinfo{journal}{Solid State Commun.} \textbf{\bibinfo{volume}{21}},
  \bibinfo{pages}{467} (\bibinfo{year}{1977}).

\bibitem[{\citenamefont{Chimenti and Clem}(1978)}]{Chimenti:1978}
\bibinfo{author}{\bibfnamefont{D.}~\bibnamefont{Chimenti}} \bibnamefont{and}
  \bibinfo{author}{\bibfnamefont{J.~R.} \bibnamefont{Clem}},
  \bibinfo{journal}{Phil. Mag.} \textbf{\bibinfo{volume}{38}},
  \bibinfo{pages}{635} (\bibinfo{year}{1978}).

\bibitem[{\citenamefont{Hurm et~al.}(1979)\citenamefont{Hurm, Selig, and
  Huebener}}]{Hurm:1979}
\bibinfo{author}{\bibfnamefont{V.}~\bibnamefont{Hurm}},
  \bibinfo{author}{\bibfnamefont{K.-P.} \bibnamefont{Selig}}, \bibnamefont{and}
  \bibinfo{author}{\bibfnamefont{R.~P.} \bibnamefont{Huebener}},
  \bibinfo{journal}{Z. Physik B} \textbf{\bibinfo{volume}{32}},
  \bibinfo{pages}{175} (\bibinfo{year}{1979}).

\bibitem[{\citenamefont{M\"{u}hlemeier
  et~al.}(1986)\citenamefont{M\"{u}hlemeier, Parisi, Huebener, and
  Buck}}]{Muhlemeier:1986}
\bibinfo{author}{\bibfnamefont{B.}~\bibnamefont{M\"{u}hlemeier}},
  \bibinfo{author}{\bibfnamefont{J.}~\bibnamefont{Parisi}},
  \bibinfo{author}{\bibfnamefont{R.}~\bibnamefont{Huebener}}, \bibnamefont{and}
  \bibinfo{author}{\bibfnamefont{W.}~\bibnamefont{Buck}}, \bibinfo{journal}{J.
  Low Temp. Phys.} \textbf{\bibinfo{volume}{64}}, \bibinfo{pages}{131}
  (\bibinfo{year}{1986}).

\bibitem[{\citenamefont{Hegger et~al.}(1999)\citenamefont{Hegger, Kantz, and
  Schreiber}}]{TISEAN}
\bibinfo{author}{\bibfnamefont{R.}~\bibnamefont{Hegger}},
  \bibinfo{author}{\bibfnamefont{H.}~\bibnamefont{Kantz}}, \bibnamefont{and}
  \bibinfo{author}{\bibfnamefont{T.}~\bibnamefont{Schreiber}},
  \bibinfo{journal}{Chaos} \textbf{\bibinfo{volume}{9}}, \bibinfo{pages}{413}
  (\bibinfo{year}{1999}).

\bibitem[{\citenamefont{Takens}(1981)}]{Takens81}
\bibinfo{author}{\bibfnamefont{F.}~\bibnamefont{Takens}},
  \bibinfo{journal}{Lect. Not. Math.} \textbf{\bibinfo{volume}{898}},
  \bibinfo{pages}{366} (\bibinfo{year}{1981}).

\bibitem[{\citenamefont{Kantz}(1994)}]{Kantz94}
\bibinfo{author}{\bibfnamefont{H.}~\bibnamefont{Kantz}},
  \bibinfo{journal}{Phys. Lett. A} \textbf{\bibinfo{volume}{185}},
  \bibinfo{pages}{77} (\bibinfo{year}{1994}).

\bibitem[{\citenamefont{Schreiber and Schmitz}(2000)}]{Schreiber00}
\bibinfo{author}{\bibfnamefont{T.}~\bibnamefont{Schreiber}} \bibnamefont{and}
  \bibinfo{author}{\bibfnamefont{A.}~\bibnamefont{Schmitz}},
  \bibinfo{journal}{Physica D} \textbf{\bibinfo{volume}{142}},
  \bibinfo{pages}{346} (\bibinfo{year}{2000}).

\bibitem[{\citenamefont{Grassberger and Procaccia}(1983)}]{Grassberger83}
\bibinfo{author}{\bibfnamefont{P.}~\bibnamefont{Grassberger}} \bibnamefont{and}
  \bibinfo{author}{\bibfnamefont{I.}~\bibnamefont{Procaccia}},
  \bibinfo{journal}{Physica D} \textbf{\bibinfo{volume}{9}},
  \bibinfo{pages}{189} (\bibinfo{year}{1983}).

\bibitem[{\citenamefont{Buck and Parisi}(1989)}]{Buck89}
\bibinfo{author}{\bibfnamefont{W.}~\bibnamefont{Buck}} \bibnamefont{and}
  \bibinfo{author}{\bibfnamefont{J.}~\bibnamefont{Parisi}},
  \bibinfo{journal}{Z. Naturforsch.} \textbf{\bibinfo{volume}{44a}},
  \bibinfo{pages}{247} (\bibinfo{year}{1989}).

\end{thebibliography}

\end{document}